\documentclass[aps,pra,twocolumn,nopacs,floatfix]{revtex4-2}
\usepackage{epsfig}
\usepackage{graphicx}
\usepackage{dcolumn}
\usepackage{amsthm,amsmath,amsfonts}
\usepackage{comment}
\usepackage[colorlinks]{hyperref}
\usepackage{xcolor}
\usepackage{mathtools}
\usepackage{orcidlink}
\usepackage{amsmath}
\usepackage{longtable}
\usepackage{booktabs}
\usepackage{ltablex}
\usepackage[normalem]{ulem}

\usepackage{supertabular}
\usepackage{adjustbox}

\begin{document}

\title{{\it Ab initio} Calculations of Electric Dipole Polarizabilities in the Li, Na and K Atoms}

\author{A. Chakraborty \orcidlink{0000-0001-6255-4584}}
\email{arupc794@gmail.com}

\author{B. K. Sahoo \orcidlink{0000-0003-4397-7965}}
\email{bijaya@prl.res.in}

\affiliation{Atomic, Molecular and Optical Physics Division, Physical Research Laboratory, Navrangpura, Ahmedabad 380009, India}

\begin{abstract}
We carry out first-principle calculations of scalar and tensor components of the static electric dipole polarizabilities of six low-lying states of lithium (Li), sodium (Na) and potassium (K) alkali atoms in the linear response approach. Results are compared from the Dirac-Hartree-Fock (DHF) method, third-order many-body perturbation theory (MBPT(3) method), random phase approximation (RPA) and singles and doubles approximated relativistic coupled-cluster theory (RCCSD method). We find the DHF and RPA results are close to each other, while the MBPT(3) and RCCSD results are close to each other. This suggests that pair-correlation effects play significant roles over core-polarization effects to determine these quantities accurately in the above alkali atoms. We also compare contributions arising through the core, core-valence and valence correlations through all the methods in Li, Na and K, which show that the core-valence contributions are negligibly small in all the methods and there is no particular trends of the core and valence correlation contributions with the size of the atom. Uncertainties to the RCCSD results are estimated to quote the final values, and they are compared with the previous calculations and experimental results.
\end{abstract}

\date{\today}

\maketitle 

\section{Introduction}\label{secin}

In the presence of an external electric field, an atomic system loses its spherical symmetricity. The distortion in the electronic charge distribution under the influence of external electric fields is quantified by electric polarizabilities \cite{Bonin1997, Haken1996, Manakov1986}. Among them, the electric dipole ($\alpha_d$) polarizability is the leading-order component and it corresponds to the second-order correction to an energy level of the atomic system for a weak electric field  \cite{Buckingham1967, Maroulis2004}. Electric polarizabilities are dependent on the angular moment components of the atomic state but independent of the field strength \cite{Bonin1997, Manakov1986}. These quantities play crucial roles in understanding interactions between an atom and external electric field. It is possible to estimate leading-order Stark shifts of energy levels for any arbitrary but sufficiently weak electric field with the knowledge of the $\alpha_d$ values of the atomic states. Accurate knowledge of $\alpha_d$ values of atomic systems are useful in estimating systematic shifts due to stray electric fields and black-body radiations in the high-precision measurements such as in atomic clocks \cite{Pal'chikov2003, Sahoo2014, Safronova2012-IEEE}, to construct effective atomic potentials to study atomic-particle collision processes \cite{Burrow1976, Jain1990, Tenfen2019}, to determine the van der Waals coefficients \cite{Tang1969, Derevianko1999, Wansbeek2008, Arora2014}, etc.. It also has applications in quantum chemistry and materials science, where it influences properties such as refractive index and dielectric constants \cite{Monin1974}. Moreover, comparison between theoretical and experimental values of $\alpha_d$ help us to understand about the behavior of electron correlations in the atomic systems \cite{Sahoo2007, Sahoo2009}. As a result, studies of $\alpha_d$ of atomic systems, particularly in the alkali atoms that are used in many high-precision experiments, has become immense research interest from both experimental and theoretical fronts \cite{Oymak2012, Peter2019}.

Lithium (Li), sodium (Na), and potassium (K) are the first three alkali metals, representing a class of elements characterized by single valence electron outside a closed-shell cores of noble gas atoms. Due to their relatively simple electronic structure both in the ground and several low-lying excited states, they serve as ideal candidates for theoretical investigations of fundamental atomic properties. Availability of lasers to access many of these energy levels helps experimentalists to perform high-precision measurements in these systems. The outer valence electron in these atoms exhibits strong sensitivity to external perturbations, including electric fields, making them highly polarizable. Accurate determination of $\alpha_d$ values of both the ground and excited states of these atoms would be useful many applications such as in guiding laser cooling and trapping techniques of atoms \cite{Ravi2012, Hall2012}, understanding underlying interactions among cold atoms \cite{Catani2006}, probing quantum phase transitions \cite{Saffman2010}, and designing quantum control experiments \cite{Joachim2000}.
 
The $\alpha_d$ values for the ground states of Li, Na and K have been measured with high precision \cite{Molof1974, Ekstrom1995, Hall1974, Miffre2006}, and theoretical calculations utilizing many-body methods are in good agreement with these experimental results \cite{Wansbeek2008, Sahoo2007, Safronova2012, Safronova1999, Nandy2012}. However, there have been limited experimental and theoretical investigations are made into the excited states of the above atoms. Additionally, the available experimental data for the excited states of K have  significant error margins \cite{Marrus1969}. This suggests the need for accurate determination of the $\alpha_d$ values of the excited states of the Li, Na and K atoms. Previously the most precise theoretical values of $\alpha_d$ for these atoms are available using the sum-over-states approach \cite{Nandy2012, Safronova2012}, which relies on dipole matrix (E1) elements and excitation energies among many bound states explicitly. Many times the E1 matrix elements are used from the calculations, while the energies are taken from the measurements to obtain $\alpha_d$ in the semi-empirical approach. Limitations of the semi-empirical approach is that it cannot include contributions from the core and high-lying (continuum) states for which first-principle calculations are needed. Again, contributions from the intermediate states with doubly-excited configurations are also mostly ignored in such approach. These limitations can be overcome by obtaining the $\alpha_d$ values in the approach similar to that was suggested by Dalgarno and Lewis \cite{Dalgarno1955}. To employ such an approach for accurate theoretical evaluations of $\alpha_d$, it is imperative to consider a potential many-body method that can account for both the electron correlation and relativistic effects rigorously.

Among the currently applied many-body methods, relativistic coupled-cluster (RCC) theory has emerged as a robust and reliable method for including both the electron correlation and leading-order relativistic effects \cite{Sahoo2025, Katyal2025}. In this work, we apply linear response approach in the RCC theory framework to determine the ground and excited states of Li, Na and K atoms.

The structure of the paper is as follows: Sec. \ref{secth} provides a brief overview of the scalar and tensor components of $\alpha_d$. In Sec. \ref{secme}, we describe various many-body methods including the RCC theory in the context of calculating $\alpha_d$. The results and their discussion are presented in Sec. \ref{secre}, where we offer detailed analyses of our theoretical findings, compare them with previous results, and highlight relevant trends along with prospects for future improvements. The paper concludes with a summary. Unless stated otherwise, all results in this work are given in atomic units (a.u.).

\begin{table}[t!]
\centering
\caption{List of $\eta_0$ and $\beta$ parameters used to define the GTOs for different symmetries to construct single-particle DHF orbitals in the present calculations.}
\begin{tabular}{p{0.7cm}p{1cm}p{1cm}p{0.8cm}p{0.8cm}p{0.8cm}p{0.8cm}p{0.8cm}}
\hline
& $s$ & $p$ & $d$ & $f$& $g$ & $h$ & $i$ \\ [1ex]
\hline
$\eta_0$ & 0.0009 & 0.0008 & 0.001 & 0.004 & 0.005 & 0.005 & 0.005 \\[1ex]
$\beta$  & 2.15 & 2.15 & 2.15 & 2.25 & 2.35 & 2.35 & 2.35 \\
\hline
\end{tabular}
\label{tab_basis}
\end{table}

\begin{table*}[t!]
\setlength{\tabcolsep}{6pt}
\caption{Calculated static scalar and tensor polarizabilities of the ground and excited states of the Li, Na and K atoms using the DHF, MBPT(3), RPA and RCCSD methods. All values are given in a.u..}
\centering
\begin{tabular}{ll cccccc cccc}
\hline\hline\\
Atom&Method & \multicolumn{6}{c}{$\alpha_d^{S}$} && \multicolumn{3}{c}{$\alpha_d^{T}$}\\
\hline\\
Li& & 2$S_{1/2}$ & 3$S_{1/2}$ & 2$P_{1/2}$ & 2$P_{3/2}$ & 3$D_{3/2}$ & 3$D_{5/2}$ && 2$P_{3/2}$ & 3$D_{3/2}$ & 3$D_{5/2}$\\
\cline{3-8}\cline{10-12}\\
&DHF   & 169.08  & 4137.48 & 136.04 & 136.06 & $-$20714.96 & $-$20724.19 && 0.02 & 15459.95 &
22100.09 \\[0.5ex]
&MBPT(3) & 164.52 & 4122.60 & 128.66 & 128.68 & $-$13679.73 & $-$13685.19 && 1.08 & 10532.99 & 15055.38 \\[0.5ex]
&RPA   & 168.46  & 4135.62 & 136.18 & 136.20 & $-$20714.64 & $-$20723.88 && $-$0.26 & 15459.62 & 22099.62\\[0.5ex]
&RCCSD & 164.14  & 4129.04 & 127.13 & 127.15 & $-$15003.92 & $-$15010.42 && 1.55 & 11458.76 & 16379.65\\[0.5ex]
\hline\\
Na & & 3$S_{1/2}$ & 4$S_{1/2}$ & 3$P_{1/2}$ & 3$P_{3/2}$ & 3$D_{3/2}$ & 3$D_{5/2}$ && 3$P_{3/2}$ & 3$D_{3/2}$ & 3$D_{5/2}$\\
\cline{3-8}\cline{10-12}\\
& DHF  & 189.36 & 3393.76 & 381.10 & 382.85 & 6149.89 & 6128.63 && $-$85.62 & $-$3358.94 & $-$4768.30 \\[0.5ex]
& MBPT(3) & 166.99 & 3138.40 & 365.69 & 367.49 & 6336.90 & 6312.28 && $-$89.76 & $-$3511.79 & $-$4981.87\\[0.5ex]
& RPA   & 187.21 & 3388.38 & 380.94 & 382.69 & 6150.04 & 6128.78 && $-$86.49 & $-$3359.62 & $-$4769.28  \\[0.5ex]
& RCCSD & 164.34  & 3116.72 & 360.96 & 362.76 & 6404.85 & 6378.82 && $-$88.20 & $-$3563.29 & $-$5053.43 \\[0.5ex]
\hline\\
K& & 4$S_{1/2}$ & 5$S_{1/2}$ & 4$P_{1/2}$ & 4$P_{3/2}$ & 3$D_{3/2}$ & 3$D_{5/2}$ && 4$P_{3/2}$ & 3$D_{3/2}$ & 3$D_{5/2}$\\
\cline{3-8}\cline{10-12}\\
& DHF & 405.90 & 5920.09 & 693.98 & 705.07 & 2147.16 & 2127.74 && $-$97.12 & $-$779.60 & $-$1089.47\\[0.5ex]
& MBPT(3) & 291.35 & 4889.27 & 629.19 & 639.99 & 1631.09 & 1614.66 && $-$130.61 & $-$589.76 & $-$821.52 \\[0.5ex]
& RPA   & 392.11 & 5889.70 & 689.55 & 700.56 & 2150.13 & 2130.72 && $-$101.66 & $-$785.06 & $-$1097.28 \\[0.5ex]
& RCCSD & 291.93 & 4945.20 & 608.42 & 619.53 & 1416.95 & 1403.40 && $-$109.37 & $-$479.99 & $-$667.61 \\[0.5ex]
\hline\hline
\end{tabular}
\label{tab1}
\end{table*}

\section{Theory} \label{secth}

A uniform static electric field is given by 
\begin{eqnarray}
\vec {\mathbb{E}}= {\cal E}_0 \hat \epsilon ,
\end{eqnarray}
where ${\cal E}_0$ is the field strength and $\hat \epsilon$ is the polarization vector. The dipole interaction of $\vec {\mathbb{E}}$ with an atom can be described by the interaction Hamiltonian
\begin{eqnarray}
H_{int}=- \vec D \cdot  \vec {\mathbb{E}} .    
\end{eqnarray}
Here $\vec D$ is the E1 operator. Since $H_{int}$ is an odd parity operator, the first-order shift to any atomic state $|J_n,M_n\rangle$, with $J_n$ and $M_n$ being the total and azimuthal angular momentum quantum numbers of states corresponding to the principal quantum number $n$, vanishes and the leading contribution comes from the second-order effects, and is expressed as
\begin{eqnarray}
\delta E(J_n,M_n)=-\frac{1}{2}\alpha_d(J_n,M_n) {\cal E}_0^2,
\end{eqnarray}
where $\alpha_d(J_n,M_n)$ depends on $J_n$ and $M_n$, and is expressed as \cite{Manakov1986, Kozlov1999, Stalnaker2006}
\begin{eqnarray}
\alpha_d(J_n,M_n)=\alpha_d^{S}(J_n) + \frac{3M_n^2-J_n(J_n+1)}{J_n(2J_n-1)}\alpha_d^{T}(J_n). 
\end{eqnarray}
Here $\alpha_d^{S}(J_n)$ and $\alpha_d^{T}(J_n)$ are called as the scalar and tensor polarizabilities, respectively. These $M_n$ independent quantities can be written in terms of reduced matrix elements as
\begin{eqnarray*}
 \alpha_d^S(J_n)&=& C_0 \sum_{k}\frac{|\langle J_n||D||J_k \rangle|^2}{E^{(0)}_{J_n} - E^{(0)}_{J_k}}
\end{eqnarray*}
and
\begin{eqnarray}
 \alpha_d^T(J_n)&=& \sum_{k} C_k \frac{|\langle J_n|| D||J_k \rangle|^2}{E^{(0)}_{J_n} - E^{(0)}_{J_k}},   
\end{eqnarray}
where $C_0 =  - \frac{2}{3(2J_n+1)}$, $C_k = \sqrt{\frac{40J_n(2J_n-1)}{3(J_n+1)(2J_n+3)(2J_n+1)}}  \times  (-1)^{J_n+J_k+1} \left\{ \begin{array}{ccc}
                    J_n& 2 & J_n\\
                  1 & J_k &1 
 \end{array}\right\} $
with the curly bracket symbol denoting the 6j coefficient, $E^{(0)}_{J_v}$ is the atomic energy value of the state with valence orbital $v$ and $J_k$ is the intermediate state. It is obvious from the angular momentum selection rule that $ \alpha_d^T(J_n)$ will be non-zero only for the states $J_n>1/2$. Following the discussions in \cite{Arora2012, Kaur2015}, we express both the scalar and tensor components as 
\begin{eqnarray}
\alpha_d^{S/T}=\alpha_{c}^{S/T}+\alpha_{cv}^{S/T}+\alpha_{v}^{S/T},
\end{eqnarray}
where subscript $c$, $cv$, and $v$ represent core, core-valence, and valence-correlation contributions, respectively. The core contribution, $\alpha_{c}^{S/T}$, corresponds to the common closed core for different valence states. In contrast, core-valence and valence contributions are defined by partitioning the intermediate state $J_k$ in the above equations into core and valence orbitals. Due to appearance of the 6j coefficient in $C_k$, the core contribution to $\alpha_d^T$ will be zero. In the sum-over-states approach, usually $\alpha_{v}^{S/T}$ is calculated using the reduced matrix elements of low-lying valence excited states and experimental energies, while contributions from $\alpha_{c}^{S/T}$ and $\alpha_{cv}^{S/T}$ are calculated using lower-order methods such as the Dirac-Hartree-Fock (DHF) method or random phase approximation (RPA) approach. 

To evaluate the $\alpha_d^{S/T}$ values by including contributions from all intermediate levels on equal footing, we express them 
\begin{eqnarray}
 \alpha_d^{S/T}&=&\langle \Psi_n^{(0)}|\tilde{D}^{S/T}|\Psi_n^{(1)} \rangle+\langle \Psi_n^{(1)}|\tilde{D}^{S/T}|\Psi_n^{(0)} \rangle\nonumber\\
 &=&2\langle \Psi_n^{(0)}|\tilde{D}^{S/T}|\Psi_n^{(1)} \rangle ,
\end{eqnarray}
where wave functions with superscript $(0)$ and $(1)$ correspond to the unperturbed atomic wave functions of the Dirac-Coulomb (DC) Hamiltonian and its first-order perturbed correction due to the E1 operator, respectively. In the above equation $\tilde{D}^S = C_0 D$ and $\tilde{D}^T = \sum_k C_k D$. In the linear response approach, we solve for the first-order perturbed wave function
\begin{eqnarray}
 (H_{DC}-E_n^{(0)})|\Psi_n^{(1)}\rangle=-D|\Psi_n^{(0)}\rangle,
\end{eqnarray}
for the DC Hamiltonian $H_{DC}$.
\begin{figure*}[t!]
\setlength{\tabcolsep}{10pt}
\centering
\begin{tabular}{c c}\\
\includegraphics[width=70mm,height=48mm]{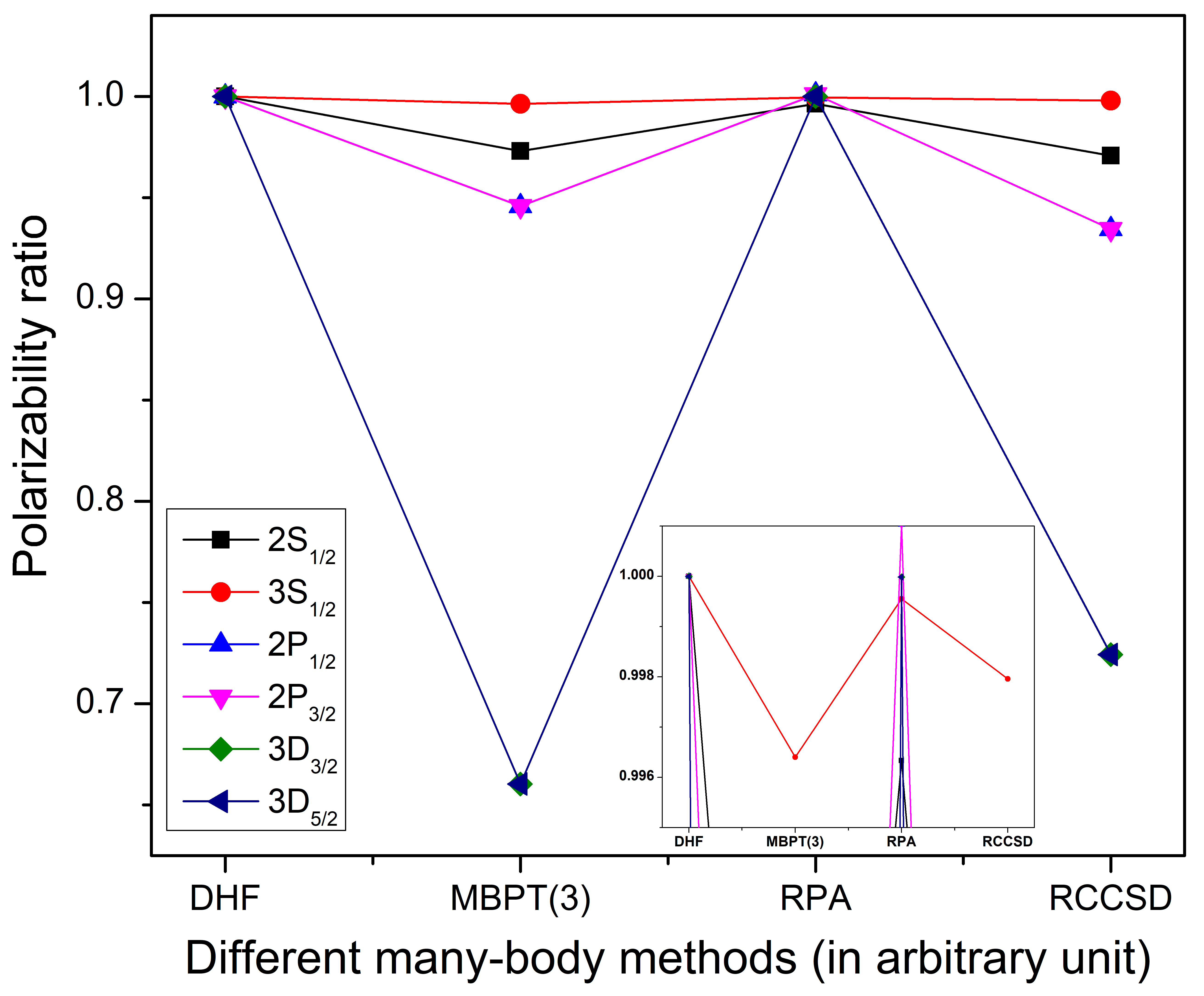} &
\includegraphics[width=70mm,height=48mm]{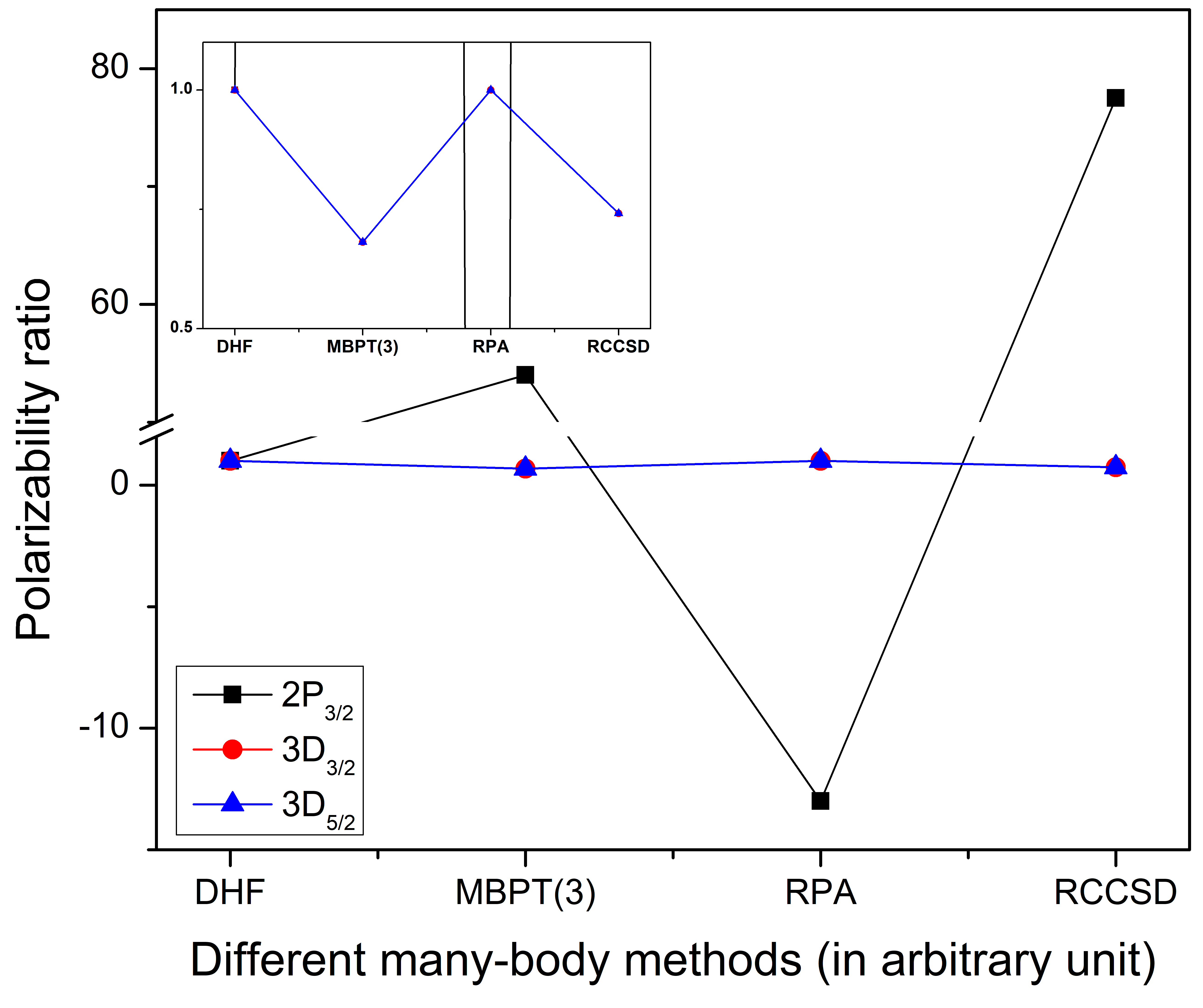}\\ 
(a) Correlation trend of $\alpha_d^S$ for Li & (b) Correlation trend of $\alpha_d^T$ for Li  \\[5ex]
\includegraphics[width=70mm,height=48mm]{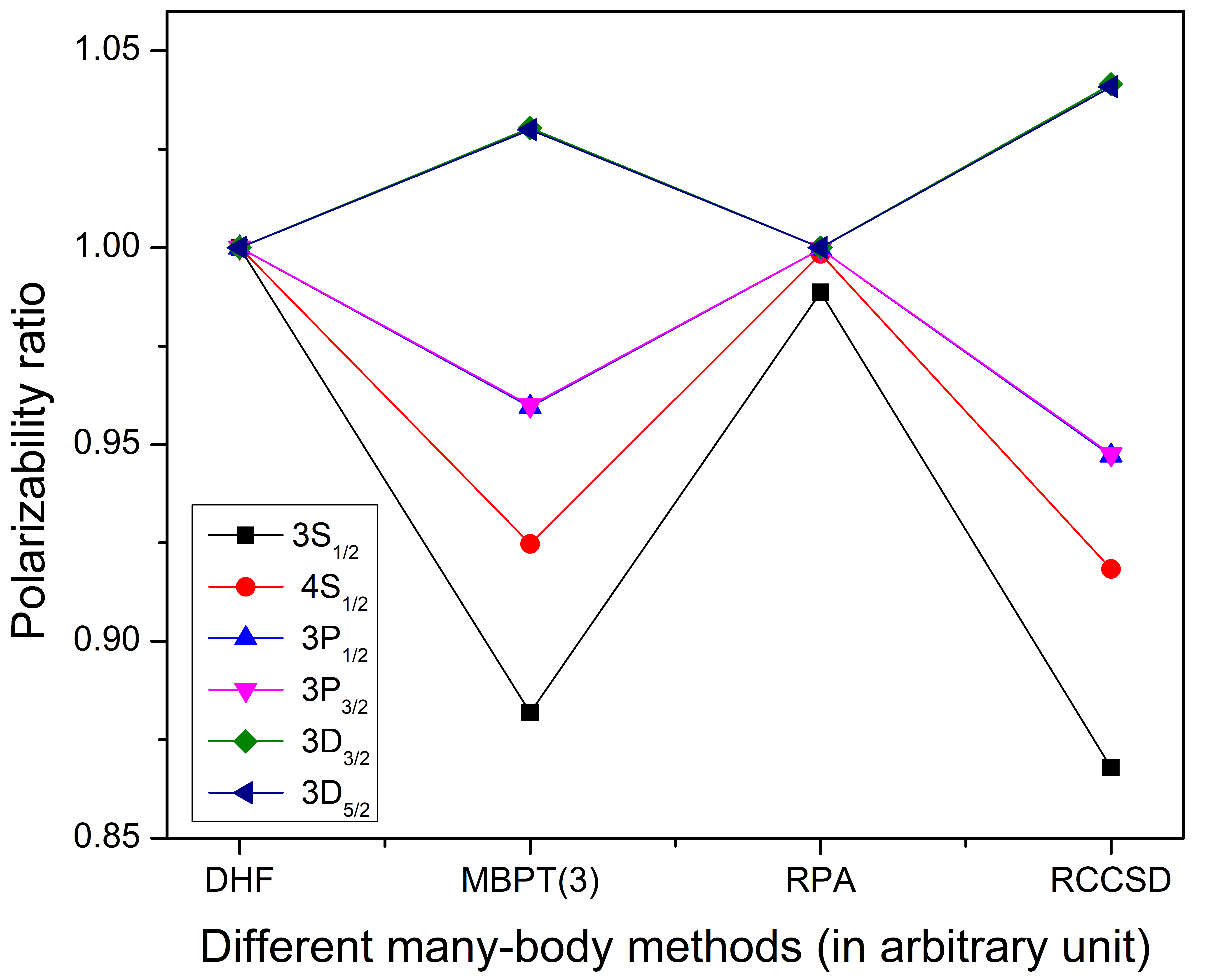} &
\includegraphics[width=70mm,height=48mm]{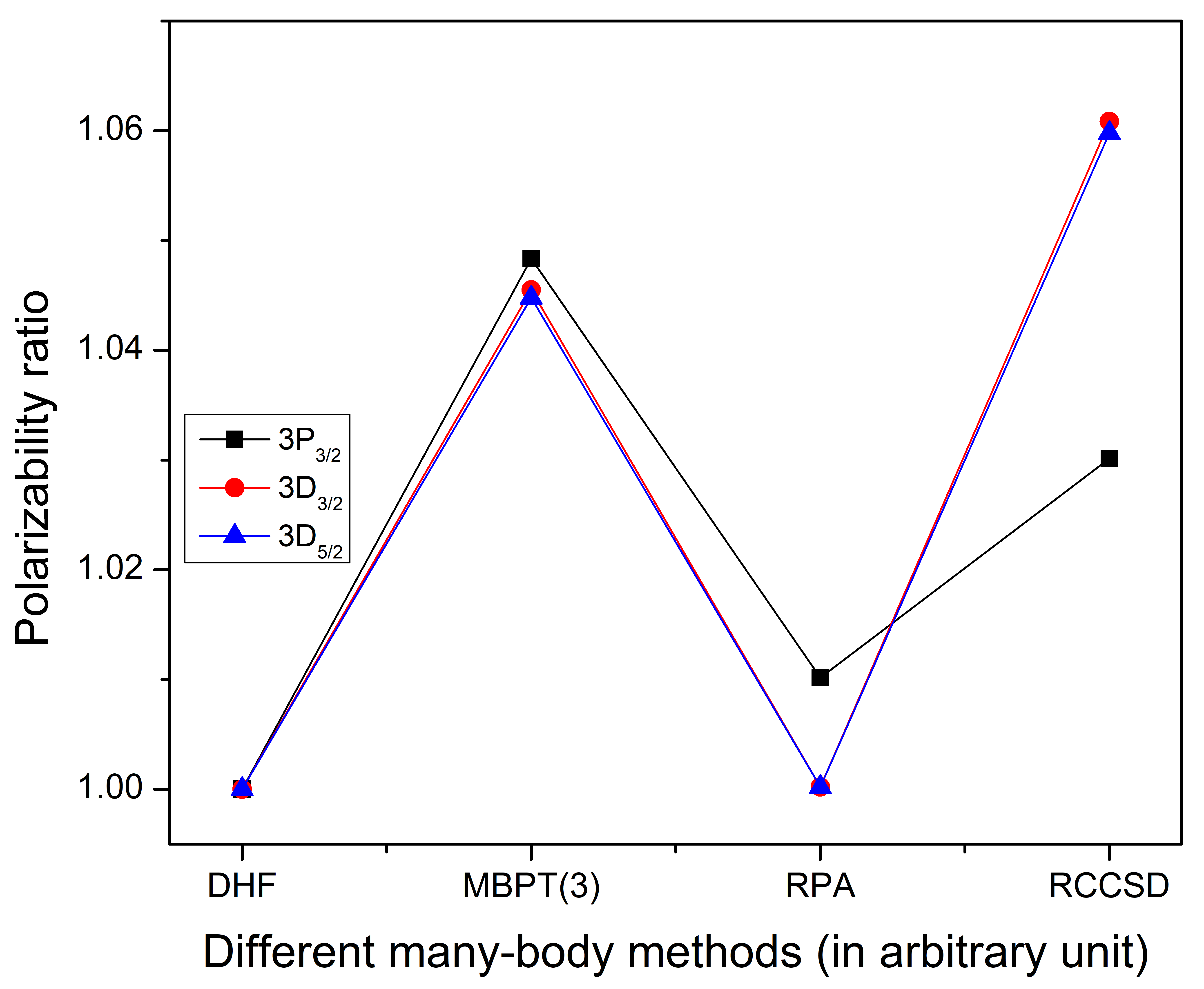}\\ 
(c) Correlation trend of $\alpha_d^S$ for Na & (d) Correlation trend of $\alpha_d^T$ for Na \\[5ex]
\includegraphics[width=70mm,height=48mm]{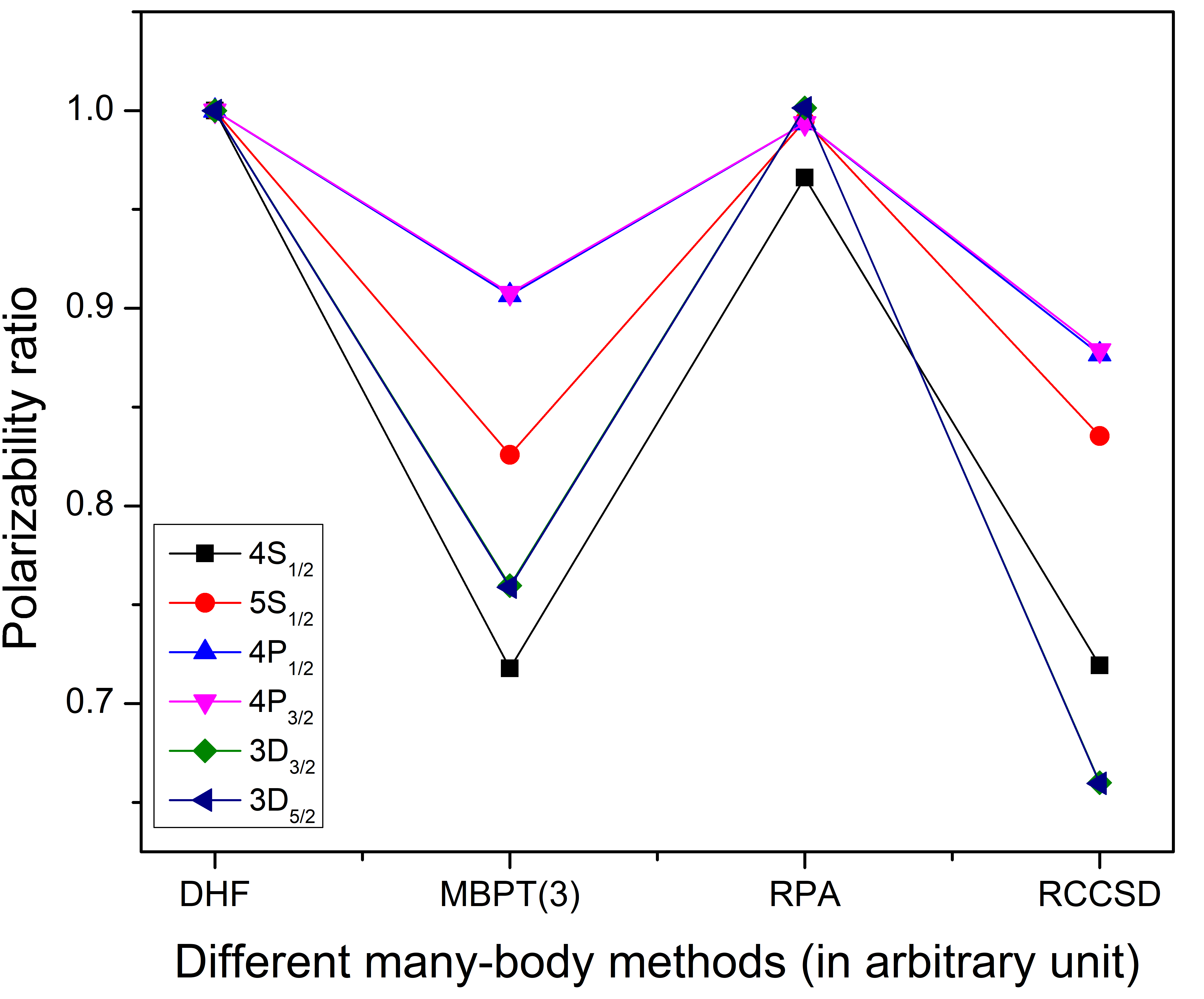} &
\includegraphics[width=70mm,height=48mm]{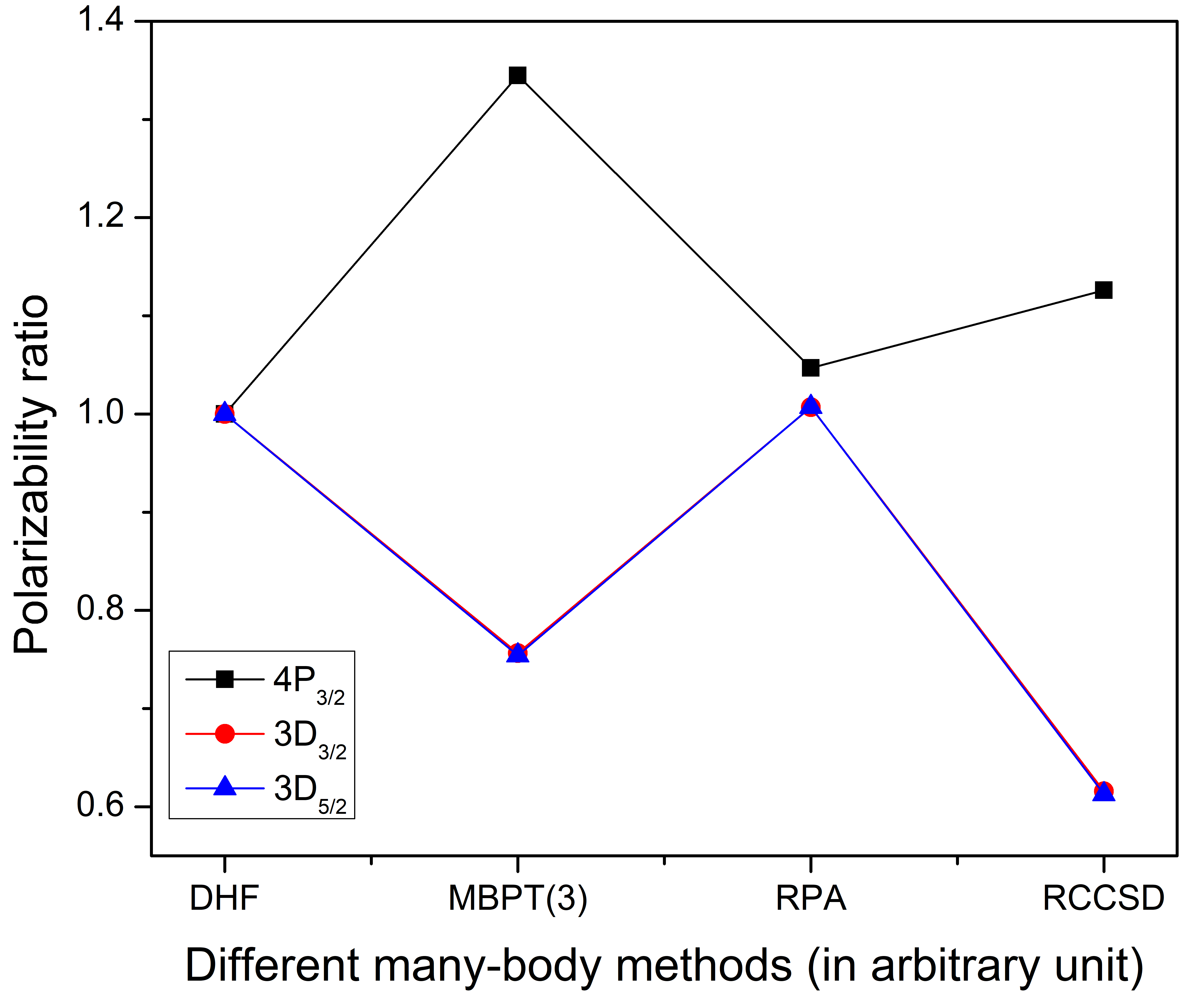}\\ 
(e) Correlation trend of $\alpha_d^S$ for K & (f) Correlation trend of $\alpha_d^T$ for K 
\end{tabular}
\caption{Ratios of scalar and tensor polarizability values from different many-body methods and their DHF values. These plots demonstrate amount of electron correlation effects captured in the many-body methods in the determination of electric dipole polarizabilities of the considered alkali atoms.}
\label{fig1}
\end{figure*} 
\begin{table*}[t]
\setlength{\tabcolsep}{3pt}
\caption{Core, core-valence, and valence contributions (in a.u.) from different methods to the $\alpha_d^S$ and $\alpha_d^T$ values of the investigated states in Li, Na and K.}
\centering
\begin{tabular}{lll cccccc cccc}
\hline\hline\\
Atom &Method&Contribution & \multicolumn{6}{c}{$\alpha_d^{S}$} && \multicolumn{3}{c}{$\alpha_d^{T}$}\\
\hline\\
Li &&& 2$S_{1/2}$ & 3$S_{1/2}$ & 2$P_{1/2}$ & 2$P_{3/2}$ & 3$D_{3/2}$ & 3$D_{5/2}$ && 2$P_{3/2}$ & 3$D_{3/2}$ & 3$D_{5/2}$\\
\cline{4-9}\cline{11-13}\\
&DHF& Core   & 0.16 &  0.16 & 0.16 & 0.16 & 0.16 & 0.16 && 0.0 & 0.0 & 0.0 \\[0.1ex]
&& Core-valence   & 0.0 & 0.0 & $\sim$0.0 & $\sim$0.0 & 0.0 & 0.0  && $\sim$0.0 & 0.0 & 0.0 \\[0.1ex]
&&Valence  & 168.92 & 4137.32 & 135.88 & 135.90 & $-20715.12$ & $-20724.35$  && 0.02 &  15459.95 &  22100.09 \\[1.1ex]

&MBPT(3)& Core & 0.19 & 0.19 & 0.19 & 0.19 & 0.19 & 0.19 && 0.0 & 0.0 & 0.0 \\[0.1ex]
&& Core-valence   & 0.0 & 0.0 & $\sim$0.0 & $\sim$0.0 & 0.0 & 0.0  && $\sim$0.0 & 0.0 & 0.0 \\[0.1ex]
&&Valence  & 164.33 & 4122.41 & 128.47 & 128.49 & $-13679.92$ & $-13685.38$ && 1.08 &  10532.99 & 15055.38 \\[1.1ex]

&RPA& Core  & 0.19 & 0.19 & 0.19 & 0.19 & 0.19 & 0.19 && 0.0 & 0.0 & 0.0 \\[0.1ex]
&& Core-valence   & 0.0 & 0.0 & $\sim$0.0 & $\sim$0.0 & 0.0 & 0.0  && $\sim$0.0 & 0.0 & 0.0 \\[0.1ex]
&&Valence    & 168.27 & 4135.43 & 135.99 & 136.01 & $-20714.83$ & $-20724.07$ && $-0.26$ & 15459.62 & 22099.62\\[1.1ex]

&RCCSD& Core    & 0.19 & 0.19 & 0.19 & 0.19 & 0.19 & 0.19 && 0.0 & 0.0 & 0.0 \\[0.1ex]
&& Core-valence   & 0.0 & 0.0 & $\sim$0.0 & $\sim$0.0 & 0.0 & 0.0  && $\sim$0.0 & 0.0 & 0.0 \\[0.1ex]
&&Valence        & 163.95 & 4128.85 & 126.94 & 126.96 & $-$15004.12 & $-$15010.61 && 1.55 & 11458.76 & 16379.65\\[1.1ex]
\hline\\
Na &&& 3$S_{1/2}$ & 4$S_{1/2}$ & 3$P_{1/2}$ & 3$P_{3/2}$ & 3$D_{3/2}$ & 3$D_{5/2}$ && 3$P_{3/2}$ & 3$D_{3/2}$ & 3$D_{5/2}$\\
\cline{4-9}\cline{11-13}\\
&DHF& Core  & 0.83 & 0.83 & 0.83 & 0.83 & 0.83 & 0.83 && 0.0 & 0.0 &  0.0 \\[0.1ex]
&& Core-valence   & $-0.02$ & $\sim$0.0 & $\sim$0.0 & $\sim$0.0 & $\sim$0.0 & $\sim$0.0 && $\sim$0.0 & $\sim$0.0 & $\sim$0.0 \\[0.1ex]
&&Valence  & 188.55 & 3392.93 & 380.27 & 382.02 & 6149.06 & 6127.80  && $-85.62$ & $-3358.94$ & $-4768.30$ \\[1.1ex]

&MBPT(3)& Core  & 0.93 & 0.93 & 0.93 & 0.93 & 0.93 & 0.93 && 0.0 & 0.0 & 0.0 \\[0.1ex]
&& Core-valence   & $-0.03$ & $\sim$0.0 & $\sim$0.0 & $\sim$0.0 & $\sim$0.0 & $\sim$0.0 && $\sim$0.0 & $\sim$0.0 & $\sim$0.0 \\[0.1ex]
&&Valence  & 166.09 & 3137.47 & 364.76 & 366.56 & 6335.97 & 6311.35 && $-89.76$ & $-3511.79$ & $-4981.87$  \\[1.1ex]

&RPA& Core  & 0.95 & 0.95 & 0.95 & 0.95 & 0.95 & 0.95 && 0.0 & 0.0 & 0.0 \\[0.1ex]
&& Core-valence & $-0.03$ & $\sim$0.0 & $\sim$0.0 & $\sim$0.0 & $\sim$0.0 & $\sim$0.0 && $\sim$0.0 & $\sim$0.0 & $\sim$0.0 \\[0.1ex]
&&Valence   & 186.29 & 3387.43 & 379.99 & 381.74 & 6149.09 & 6127.83 && $-86.49$ & $-3359.62$ & $-4769.28$ \\[1.1ex]

&RCCSD& Core  &  1.00 & 1.00 & 1.00 & 1.00 & 1.00 & 1.00 && 0.0 & 0.0 & 0.0 \\[0.1ex]
&& Core-valence   & $-$0.03 & $-$0.01 & $\sim$0.0 & $\sim$0.0 & $\sim$0.0 & $\sim$0.0 && $\sim$0.0 & $\sim$0.0 & $\sim$0.0 \\[0.1ex]
&& Valence   & 163.37 & 3115.72 & 359.96 & 361.76 & 6403.85 & 6377.82 && $-$88.20 & $-$3563.29 & $-$5053.43 \\[1.1ex]
\hline\\
K &&& 4$S_{1/2}$ & 5$S_{1/2}$ & 4$P_{1/2}$ & 4$P_{3/2}$ & 3$D_{3/2}$ & 3$D_{5/2}$ && 4$P_{3/2}$ & 3$D_{3/2}$ & 3$D_{5/2}$\\
\cline{4-9}\cline{11-13}\\
&DHF& Core & 5.47 & 5.47 & 5.47 & 5.47 & 5.47 & 5.47 && 0.0 & 0.0 & 0.0 \\[0.1ex]
&& Core-valence  & $-0.12$ & $-0.02$ & $\sim0.0$ & $\sim0.0$ & $-0.03$ & $-0.03$ && $\sim0.0$ & 0.02 & 0.03 \\[0.1ex]
&&Valence    & 400.55 & 5914.64 & 688.51 & 699.60 & 2141.72 &  2122.30 && $-97.12$ & $-779.62$ & $-1089.50$ \\[1.1ex]

&MBPT(3)& Core  & 4.47 & 4.47 & 4.47 & 4.47 & 4.47 & 4.47 && 0.0 & 0.0 & 0.0 \\[0.1ex]
&& Core-valence & $-0.20$ & $-0.04$ & $\sim0.0$ & $\sim0.0$ & $-0.03$ & $-0.03$ && $\sim0.0$ & $0.02$ & $0.03$ \\[0.1ex]
&&Valence  & 287.08 & 4884.84 & 624.72 & 635.52 & 1626.65 & 1610.22 && $-130.61$ & $-589.78$ & $-821.55$ \\[1.1ex]

&RPA& Core  & 5.46 & 5.46 & 5.46 & 5.46 & 5.46 & 5.46 && 0.0 & 0.0 & 0.0 \\[0.1ex]
&& Core-valence   & $-0.18$ & $-0.03$ & $\sim0.0$ & $\sim0.0$ &  $-0.03$ &  $-0.03$ && $\sim0.0$ & 0.02 & 0.03 \\[0.1ex]
&&Valence   & 386.83 & 5884.27 & 684.09 & 695.10 & 2144.70 & 2125.29 && $-101.66$ & $-785.08$ & $-1097.31$ \\[1.1ex]

&RCCSD&Core  & 5.56 & 5.56 & 5.56 & 5.56 & 5.56 & 5.56 && 0.0 & 0.0 & 0.0 \\[0.1ex]
&& Core-Valence & $-$0.17 & $-$0.03 & $\sim0.0$ & $\sim0.0$ & $-$0.02 & $-$0.02 && $\sim0.0$ & 0.02 & 0.02 \\[0.1ex]
&& Valence   & 286.54 & 4939.67 & 602.86 & 613.97 & 1411.42 & 1397.86 && $-$109.37 & $-$480.01 & $-$667.64 \\[1.1ex]
\hline\hline
\end{tabular}
\label{tab2}
\end{table*}
\begin{table*}[]
\setlength{\tabcolsep}{5pt}
\centering
\caption{Contributions from different RCC terms to the polarizability values of $\alpha_d^S$ and $\alpha_d^T$ of all the states of Li, Na, and K atoms considered here. Terms with subscript $`c$' and $`cv$' correspond to core and core-valence contributions respectively. Contributions given under `Norm' represent corrections to the results due to normalization factors of the wave functions. Contributions from other non-linear terms of the RCCSD method are given together under ``Others".}
\begin{tabular}{ll cccccc cccc}
\hline\hline
Atom&Term & \multicolumn{6}{c}{$\alpha_d^{S}$} && \multicolumn{3}{c}{$\alpha_d^{T}$}\\
\hline
Li & & 2$S_{1/2}$ & 3$S_{1/2}$ & 2$P_{1/2}$ & 2$P_{3/2}$ & 3$D_{3/2}$ & 3$D_{5/2}$ && 2$P_{3/2}$ & 3$D_{3/2}$ & 3$D_{5/2}$\\
\cline{3-8}\cline{10-12}\\
& $(\bar{\tilde{D}}T_1^{(1)})_{c}$ & 0.10 & 0.10 & 0.10 & 0.10 & 0.10 & 0.10 && 0.0 & 0.0 & 0.0\\[0.5ex]
& $(T_1^{(1)\dagger}\bar{\tilde{D}})_{c}$ & 0.10 & 0.10 & 0.10 & 0.10 & 0.10 & 0.10 && 0.0 & 0.0 & 0.0\\[0.5ex]
& $(\bar{\tilde{D}}T_1^{(1)})_{cv}$ & 0.0 & 0.0 & $\sim$0.0 & $\sim$0.0 & 0.0 & 0.0 && $\sim$0.0 & 0.0 & 0.0\\[0.5ex]
& $(T_1^{(1)\dagger}\bar{\tilde{D}})_{cv}$ & 0.0 & 0.0 & $\sim$0.0 & $\sim$0.0 & 0.0 & 0.0 && $\sim$0.0 & 0.0 & 0.0\\[0.5ex]
&  $\bar{\tilde{D}}S_{1v}^{(1)}$  & 82.89 & 2079.11 & 65.27 & 65.28 & $-$7512.71 & $-$7515.96 && 0.21 & 5737.62 & 8201.60\\[0.5ex]
& $S_{1v}^{(1)\dagger} \bar{\tilde{D}}$ & 82.89  & 2079.11 & 65.27 & 65.28 & $-$7512.71 & $-$7515.96 && 0.21 &  5737.62 & 8201.60\\[0.5ex]
& $\bar{\tilde{D}}S_{2v}^{(1)}$  & $-0.18$  & $-$0.53 & 0.01 & 0.01 & 0.04 & 0.04 && $-$0.05 & $-$0.06 & $-$0.09\\[0.5ex]
& $S_{2v}^{(1)\dagger}\bar{\tilde{D}}$ & $-0.18$  & $-$0.53 & 0.01 & 0.01 & 0.04 & 0.04 && $-$0.05 & $-$0.06 & $-$0.09\\[0.5ex]
& Norm & $-$0.08  & $-$0.90 & $-$0.09 & $-$0.09 & 0.28 & 0.28 && $\sim$0.0 & $-0.21$ & $-$0.31\\[0.5ex]
&  Others & $-$1.40  & $-$27.42 & -3.54 & 3.54 & 20.94 & 20.94 && 1.23 & $-$16.15 & $-$23.06\\[0.5ex]
\hline
Na & & 3$S_{1/2}$ & 4$S_{1/2}$ & 3$P_{1/2}$ & 3$P_{3/2}$ & 3$D_{3/2}$ & 3$D_{5/2}$ && 3$P_{3/2}$ & 3$D_{3/2}$ & 3$D_{5/2}$\\
\cline{3-8}\cline{10-12}\\
& $(\bar{\tilde{D}}T_1^{(1)})_{c}$& 0.49  & 0.49 & 0.49 & 0.49 & 0.49 & 0.49 && 0.0 & 0.0 & 0.0\\[0.5ex]
& ($T_1^{(1)\dagger} \bar{\tilde{D}})_{c}$ & 0.49  & 0.49 & 0.49 & 0.49 & 0.49 & 0.49 && 0.0 & 0.0 & 0.0\\[0.5ex]
& $(\bar{\tilde{D}}T_1^{(1)})_{cv}$ & $-0.02$  & $-0.01$ & $\sim$0.0 & $\sim$0.0 & $\sim$0.0 & $\sim$0.0 && $\sim$0.0 & $\sim$0.0 & $\sim$0.0\\[0.5ex]
& ($T_1^{(1)\dagger} \bar{\tilde{D}})_{cv}$ & $-0.02$  & $-0.01$ & $\sim$0.0 & $\sim$0.0 & $\sim$0.0 & $\sim$0.0 && $\sim$0.0 & $\sim$0.0 & $\sim$0.0\\[0.5ex]
& $\bar{\tilde{D}}S_{1v}^{(1)}$ & 85.30 & 1617.34 & 186.03 & 186.93 & 3235.13 & 3222.00 && $-$46.48 & $-$1800.83 & $-$2553.96\\[0.5ex]
& $S_{1v}^{(1)\dagger} \bar{\tilde{D}}$ & 85.30 & 1617.34 & 186.03 & 186.93 & 3235.13 & 3222.00 && $-$46.48 & $-$1800.83 & $-$2553.96\\[0.5ex]
& $\bar{\tilde{D}}S_{2v}^{(1)}$ & $-$0.58 & $-$1.21 & $-$0.08 & $-$0.08 & $-$0.05 & $-$0.05 && $-$0.19 & $-$0.12 & $-$0.17\\[0.5ex]
& $S_{2v}^{(1)\dagger}\bar{\tilde{D}}$ & $-$0.58 & $-$1.21 & $-$0.08 & $-$0.08 & $-$0.05 & $-$0.05 && $-$0.19 & $-$0.12 & $-$0.17\\[0.5ex]
& Norm & $-$0.57 & $-$6.98 & $-$0.47 & $-$0.47 & $-$1.31 & $-$1.30 && 0.11 & 0.73 & 1.03\\[0.5ex]
&  Others & $-$5.47 & $-$109.52 & $-$11.44 & $-$11.44 & $-$64.98 & $-$64.76 && 5.03 & 37.88 & 53.80\\[0.5ex]
\hline
K & & 4$S_{1/2}$ & 5$S_{1/2}$ & 4$P_{1/2}$ & 4$P_{3/2}$ & 3$D_{3/2}$ & 3$D_{5/2}$ && 4$P_{3/2}$ & 3$D_{3/2}$ & 3$D_{5/2}$\\
\cline{3-8}\cline{10-12}\\
& $(\bar{\tilde{D}}T_1^{(1)})_{c}$ & 2.69 & 2.69 & 2.69 & 2.69 & 2.69 & 2.69 && 0.0 & 0.0 & 0.0 \\[0.5ex]
& ($T_1^{(1)\dagger} \bar{\tilde{D}})_{c}$ & 2.69 & 2.69 & 2.69 & 2.69 & 2.69 & 2.69 && 0.0 & 0.0 & 0.0 \\[0.5ex]
& $(\bar{\tilde{D}}T_1^{(1)})_{cv}$ & $-0.09$ & $-0.02$ & $\sim$0.0 & $\sim$0.0 & $-0.01$ & $-0.01$ && $\sim$0.0 & 0.01 & 0.01 \\[0.5ex]
& ($T_1^{(1)\dagger} \bar{\tilde{D}})_{cv}$& $-0.09$ & $-0.02$ & $\sim$0.0 & $\sim$0.0 & $-0.01$ & $-0.01$ && $\sim$0.0 & 0.01 & 0.01 \\[0.5ex]
&  $\bar{\tilde{D}}S_{1v}^{(1)}$  & 158.89  & 2685.57 & 326.45 & 332.13 & 849.42 & 841.41 && $-$63.32 & $-$305.63 & $-$426.16 \\[0.5ex]
& $S_{1v}^{(1)\dagger} \bar{\tilde{D}}$ & 158.89  & 2685.57 & 326.45 & 332.13 & 849.42 & 841.41 && $-$63.32 & $-$305.63 & $-$426.16 \\[0.5ex]
& $\bar{\tilde{D}}S_{2v}^{(1)}$  & $-$2.46  & $-$6.22 & $-1.29$ & $-$1.31 & 0.26 & 0.26 && $-0.74$ & $-$0.83 & $-$1.19\\[0.5ex]
& $S_{2v}^{(1)\dagger}\bar{\tilde{D}}$ & $-$2.46  & $-$6.22 & $-1.29$ & $-$1.31 & 0.26 & 0.26 && $-0.74$ & $-$0.83 & $-$1.19\\[0.5ex]
& Norm & $-$3.72  & $-$55.91 & $-$4.13 & $-$4.14 & $-$27.87 & $-$27.46 && 0.74 & 9.48 & 13.11\\[0.5ex]
&  Others & $-$22.41  & $-$362.93 & $-$43.15 & $-$43.35 & $-$259.90 & $-$257.84 && 18.01 & 123.43 & 173.96\\[0.5ex]
\hline\hline
\end{tabular}
\label{tab3}
\end{table*}

\section{Methodology} \label{secme}

The DC Hamiltonian $H_{DC}$ for atomic systems is given in a.u. by
\begin{eqnarray}\label{eq:DC}
H_{DC} &=& \sum_i \left [c {\vec \alpha}_i^D \cdot {\vec p}_i+(\beta_i^D-1)c^2+V_n(r_i)\right ] +\sum_{i,j>i}\frac{1}{r_{ij}}, \nonumber 
\end{eqnarray}
where $c$ is speed of light, $\alpha^D$ and $\beta^D$ are the Dirac matrices, ${\vec p}$ is the single particle momentum operator, $V_n(r)$ denotes nuclear potential seen by an electron and $\frac{1}{r_{ij}}$ represents the Coulomb potential between the electrons. 

In order to obtain final unperturbed wave functions, we first determine the DHF wave function, $|\Phi_0 \rangle$, due to $H_{DC}$ of the closed-shell [$np^6$] core. In the next step, the exact atomic wave function of the closed-core, $|\Psi_0^{(0)} \rangle$ is determined by incorporating the electron correlation effects due to the residual Coulomb interactions, $V_{res}=H_{DC}-H_{DHF}$ with the DHF Hamiltonian $H_{DHF}$. We define wave operator $\Omega_0^{(0)}$, such that $|\Psi_0^{(0)}\rangle=\Omega_0^{(0)}|\Phi_0\rangle$. In the final step, we obtain the intended wave functions by appending the required valence orbital, $v$, of the state to the closed-core configuration. For this purpose, the modified DHF wave function is defined as $|\Phi_v \rangle = a_v^{\dagger} |\Phi_0 \rangle$. The final unperturbed wave function can be defined as $|\Psi_v^{(0)}\rangle=(\Omega_0^{(0)}+\Omega_v^{(0)})|\Phi_v\rangle$. Here $\Omega_0^{(0)}$ includes correlations only from the closed-core where the $\Omega_v^{(0)}$ captures the correlation effects from all orbitals including valence electron. In the similar fashion we define two wave operators, $\Omega_0^{(1)}$ and $\Omega_v^{(1)}$, to obtain the first-order perturbed wave functions due to the E1 operator
\begin{eqnarray}
|\Psi_0^{(1)} \rangle = \Omega_0^{(1)} |\Phi_0 \rangle 
\end{eqnarray}
and
\begin{eqnarray}
|\Psi_v^{(1)} \rangle = (\Omega_0^{(1)} +  \Omega_v^{(1)}) |\Phi_v \rangle .  
\end{eqnarray}

The DHF expression for $\alpha_d^{S/T}$ can be given by 
\begin{eqnarray}
\alpha_d^{S/T}  &=& 2\big[\langle \Phi_v | \Omega_0^{DHF,(0) \dagger} \tilde{D} \Omega_0^{DHF,(1) }| \Phi_v \rangle \nonumber \\
&& + \langle \Phi_v | \Omega_v^{DHF,(0) \dagger} \tilde{D} \Omega_v^{DHF,(1) } | \Phi_v \rangle\big],
\label{eqnhf}
\end{eqnarray}
where $\Omega_0^{DHF,(0) } = \Omega_v^{DHF,(0) } =1$, $\Omega_0^{DHF,(1)} = \sum_{ap} \frac{ \langle \Phi_a^p | D | \Phi_0 \rangle} { \epsilon_p - \epsilon_a} a_p^{\dagger} a_a$ and $\Omega_v^{DHF,(1)} = \sum_p \frac{ \langle \Phi_v^p | D | \Phi_v \rangle} { \epsilon_p - \epsilon_v} a_p^{\dagger} a_v$ with $\epsilon_i$ is energy of the the $i^{th}$ DHF orbital. Here, $a,b$ denote for core orbitals, $p,q$ denote for virtual orbitals and $|\Phi_{ab\cdots}^{pq\cdots} \rangle = a_p^{\dagger} a_q^{\dagger} \cdots a_b a_a |\Phi_0 \rangle$. 

To understand importance of contributions from $V_{res}$ to $\alpha_d^{S/T}$, we include the correlation effects first at the many-body perturbation (MBPT) theory. In MBPT method the amplitudes of the unperturbed and perturbed wave operators are estimated using the Bloch equation \cite{Lindgren1985, Singh2014, Chakraborty2023-2}. The Block equation for the unperturbed case are as follows \cite{Lindgren1985}
\begin{eqnarray*}
\left[\Omega^{MBPT,(0)}_0,H_{DHF}\right]=\big(V_{res}\Omega^{MBPT,(0)}_0\big)_{l}
\end{eqnarray*}
and
\begin{eqnarray}
\left[\Omega_v^{MBPT,(0)}, H_{DHF}\right]&=&\big[V_{res} (\Omega^{MBPT,(0)}_0+\Omega^{MBPT,(0)}_v)\nonumber\\
&& -\Omega^{MBPT,(0)}_v (V_{res}
(\Omega^{MBPT,(0)}_0\nonumber\\
&&+\Omega^{MBPT,(0)}_v))\big]_{l} , 
\end{eqnarray}
where `$l$' means that only the linked diagrams will contribute to the wave operator. The Bloch's equations for the first-order perturbed wave operators are given by   \cite{Singh2014, Sahoo2017, Chakraborty2023-2}.
\begin{eqnarray*}
 [\Omega_0^{MBPT,(1)}, H_{DHF}] &=& (D \Omega_0^{MBPT,(0)} +V_{res} \Omega_0^{MBPT,(1)} )_l  
\end{eqnarray*}
and
\begin{eqnarray*}
 [\Omega_v^{MBPT,(1)}, H_{DHF}] &=& \big(D (\Omega_0^{MBPT,(0)} + \Omega_v^{MBPT,(0)}) \nonumber\\
 &&+ V_{res} ( \Omega_0^{MBPT,(1)} +\Omega_v^{MBPT,(1)} )\big)_l\nonumber \\
  &&  - \Omega_v^{MBPT,(1)} E_v^{(0)}.  
\end{eqnarray*}
The energy, $E^{(0)}_v$, of the state is estimated by
\begin{eqnarray}
E^{(0)}_v=\langle \Phi_v | H_{DC} (\Omega_0^{(0)} +\Omega_v^{(0)}) |\Phi_v \rangle .
\end{eqnarray}
In the MBPT(n) method with $n$ representing n-order in perturbation, we express polarizability expression as
\begin{eqnarray}
\alpha_d^{S/T}  &=& \frac{2}{N}\big[\langle \Phi_v | \Omega_0^{MBPT,(0) \dagger} \tilde{D} \Omega_0^{MBPT,(1) }| \Phi_v \rangle \nonumber \\
&& + \langle \Phi_v | \Omega_v^{MBPT,(0) \dagger} \tilde{D} \Omega_v^{MBPT,(1) } | \Phi_v \rangle\big],
\label{eqnMBPT}
\end{eqnarray}
where $N$ is the normalization constant. In this work, we restrict ourselves up to MBPT(3) method.

After considering $V_{res}$ in the MBPT(3) method, we use the RPA method to understand roles of core-polarization (CP) effects to all-order in the evaluations of $\alpha_d^{S/T}$ by expressing
\begin{eqnarray}
\alpha_d^{S/T}  &=& 2\big[\langle \Phi_v | \Omega_0^{RPA,(0) \dagger} \tilde{D} \Omega_0^{RPA,(1) }| \Phi_v \rangle \nonumber \\
&& + \langle \Phi_v | \Omega_v^{RPA,(0) \dagger} \tilde{D} \Omega_v^{RPA,(1) } | \Phi_v \rangle\big],
\label{eqnRPA1}
\end{eqnarray}
where $\Omega_{0/v}^{RPA,(0)}\equiv\Omega_{0/v}^{DHF,(0)}$ and the amplitudes of the first-order RPA wave operators are obtained by 
\begin{eqnarray}
(H_{DHF} - E^{(0)}_0) \Omega_0^{RPA,(1)} |\Phi_0 \rangle =  - D | \Phi_0 \rangle - U_{D}^{(1)}  |\Phi_0 \rangle  \nonumber 
\end{eqnarray}
and
\begin{eqnarray}
(H_{DHF} - E^{(0)}_v) \Omega_v^{RPA,(1) } |\Phi_v \rangle =  - D | \Phi_v \rangle - U_{D}^{(1)}  |\Phi_v \rangle . \nonumber  
\end{eqnarray}
Here $E^{(0)}_v= E^{(0)}_0 + \epsilon_v$ with $E^{(0)}_0=\sum_a^{N_c}\epsilon_a$ and $N_c$ denotes number of occupied orbitals in the system. $U_{D}^{(1)}$ is the perturbed DHF potential and is defined as
\begin{eqnarray}
  U_{D}^{(1)} | \Phi_i \rangle = \sum_b^{N_c} \left [ \langle b | V_{res} \Omega_i^{RPA,(1)} |b \rangle |i \rangle \right. \nonumber  \\ 
  \left.  -  \langle b | V_{res} \Omega_i^{RPA,(1)}  |i \rangle |b \rangle \right. \nonumber  \\ 
  \left. + \langle b | \Omega_i^{RPA,(1)\dagger} V_{res}  | b \rangle |i \rangle \right. \nonumber  \\ 
  \left. -  \langle b |\Omega_i^{RPA,(1)\dagger} V_{res}  |i \rangle |b \rangle  \right ].  
\label{eqhfu1c}
\end{eqnarray}
As can be observed, correlation effects in the unperturbed wave functions are missing in the RPA. To account for the correlation effects in the unperturbed wave functions, contributions from the pair correlation (PC) effects to all-orders and correlations among the CP and PC correlations, we use the RCC theory.

In the RCC theory {\it ansatz}, the unperturbed wave operators are given by \cite{Lindgren1985, Mukherjee1979}
\begin{eqnarray}
\Omega_0^{RCC,(0)} = e^{T^{(0)}}  
\end{eqnarray}
and
\begin{eqnarray}
\Omega_v^{RCC,(0)} = e^{T^{(0)}} S_v^{(0)}.
\label{eqrcc}
\end{eqnarray}
Extending these definitions to the first-order perturbed wave functions, we can define the corresponding wave operators as 
\begin{eqnarray}
\Omega_0^{RCC,(1)} = e^{T^{(0)}} T^{(1)} 
\end{eqnarray}
and
\begin{eqnarray}
\Omega_v^{RCC,(1)} = e^{T^{(0)}} \left ( S_v^{(1)} + S_v^{(0)}T^{(1)} \right ).
\end{eqnarray}

In the present work, we consider all possible single and double excitations in the RCC theory (RCCSD method). The singles and doubles excited RCC wave operators are denoted by additional subscripts 1 and 2, respectively. Thus, we define in the RCCSD method
\begin{eqnarray}
T^{(0)} &=&  T_{1}^{(0)} + T_{2}^{(0)} , \nonumber \\
T^{(1)} &=&  T_{1}^{(1)} + T_{2}^{(1)}, \nonumber \\
S_v^{(0)} &=&  S_{1v}^{(0)} + S_{2v}^{(0)} \nonumber 
\end{eqnarray}
and
\begin{eqnarray}
S_v^{(1)} &=&  S_{1v}^{(1)} + S_{2v}^{(1)}.
\end{eqnarray}

The amplitude determining equations for the unperturbed wave operator in the RCC theory are given by
\begin{eqnarray}
\langle\Phi^*_0|(H_{DC}e^{T^{(0)}})_{l}|\Phi_0\rangle=0
\end{eqnarray}
and 
\begin{eqnarray}
\langle\Phi^*_v|(H_{DC}e^{T^{(0)}})_{op}\{1+S_v^{(0)}\}|\Phi_v\rangle=E_v^{(0)} \langle\Phi^*_v|S_v^{(0)}|\Phi_v\rangle, \nonumber\\
\end{eqnarray}
where `$op$' corresponds to the open part in $H_{DC}e^{T^{(0)}}$ contraction. `*' denotes excited states with respect to $|\Phi_0\rangle$ or $|\Phi_v\rangle$ configuration. The first-order perturbed RCC operator amplitudes are determined as \cite{Sahoo2007} 
\begin{eqnarray}
 \langle \Phi_0^* | (H_{DC}e^{T^{(0)}})_{l} T^{(1)} | \Phi_o \rangle = -  \langle \Phi_0^* | (D e^{T^{(0)}})_{l} | \Phi_0 \rangle 
\end{eqnarray}
and 
\begin{eqnarray}
 \langle \Phi_v^* |  [(H_{DC} e^{T^{(0)}})_{l} - E_v^{(0)}] S_v^{(1)} | \Phi_v \rangle = -\langle \Phi_v^* | [  (D e^{T^{(0)}})_{l} && \nonumber \\  + (H_{DC}e^{T^{(0)}})_{l} T^{(1)} ] \{ 1+S_v^{(0)} \}  | \Phi_v \rangle.  \ \ \ \ 
 \label{eqsv}
\end{eqnarray}
So in the RCC theory the polarizability expression takes the form
\begin{eqnarray}
\alpha_d^{S/T} =2\frac{\langle \Phi_v |\{ 1+ S_v^{(0)} \}^{\dagger} \bar{\tilde{D}} \{T^{(1)}(1+ S_v^{(0)}) + S_v^{(1)}\} |\Phi_v \rangle}{\langle \Phi_v | \{S_v^{(0)\dagger} +1 \} \bar{N} \{ 1+ S_v^{(0)} \} |\Phi_v \rangle} . \ \ \ \ \ \ \ 
\label{e1pol}
\end{eqnarray}
In the above expression $\bar{\tilde{D}}=e^{T^{(0)\dagger}}\tilde{D}e^{T^{(0)}}$ and $\bar{N}=e^{T^{(0)\dagger}}e^{T^{(0)}}$.
\begin{table*}[t]
\setlength{\tabcolsep}{1pt}
\centering
\caption{List of recommended values for $\alpha_d^S$ and $\alpha_d^T$ (in a.u.) from our calculations. We also compare our results with recent high-precision relativistic calculations and experimental data.}
\begin{tabular}{ll rcr crcc}
\hline\hline
Atom & State & \multicolumn{3}{c}{$\alpha_d^S$} && \multicolumn{3}{c}{$\alpha_d^T$} \\
\cline{3-5}\cline{7-9}\\
& & This work & Theory & Experiment && This work & Theory & Experiment\\
\hline
Li&2$S_{1/2}$&164.14(22)&164.23 \cite{Wansbeek2008}, 162.48(56) \cite{Sahoo2007}&164(3.4) \cite{Molof1974} &&  &&\\
&&&164.16(5) \cite{Safronova2012}, 164.11(3) \cite{Tang2010} &164.20(2) \cite{Miffre2006}&&  &&\\
&&&164.19 \cite{Tang2014}&  &&  &&\\[2ex]   
&3$S_{1/2}$&4129.0(8.4)&4115.73 \cite{Wansbeek2008}, 4130(1) \cite{Safronova2012}&  &&  &&\\
&&&4132.7 \cite{Tang2014}&  &&  &&\\[2ex]   
&2$P_{1/2}$&127.13(48)&127.15 \cite{Wansbeek2008}, 126.97(5) \cite{Safronova2012}&126.9(3) \cite{Windholz1992} &&  &&\\
&&&126.97 \cite{Tang2014}&127(3.4) \cite{Hunter1991} &&  &&\\[2ex]
&2$P_{3/2}$&127.15(49)&127.09 \cite{Wansbeek2008}, 126.98(5) \cite{Safronova2012}&126.9(3) \cite{Windholz1992} &&1.55(42)&1.597 \cite{Wansbeek2008}, 1.612(4) \cite{Safronova2012}&1.64(4) \cite{Windholz1992}\\
 &&& 126.99 \cite{Tang2014}&  && & 1.6333 \cite{Tang2014}&\\[2ex]
&3$D_{3/2}$&$-$15003(16)&$-$14820 \cite{Wansbeek2008}, $-$14925(8) \cite{Safronova2012}&  &&11458(15)&11460 \cite{Wansbeek2008}, 11405(6) \cite{Safronova2012}& \\
&&&$-$14914 \cite{Tang2014}&  &&  & 11399 \cite{Tang2014}&\\[2ex]
&3$D_{5/2}$&$-$15010(15)&$-$14930 \cite{Wansbeek2008}, $-$14928(9) \cite{Safronova2012}&  &&16379(15)&16290 \cite{Wansbeek2008}, 16297(7) \cite{Safronova2012}&\\
&&&$-$14933 \cite{Tang2014}&  &&  & 16303 \cite{Tang2014}&\\[2ex]
\hline   
Na&3$S_{1/2}$&164.3(1.3)& 163.78(48) \cite{Sahoo2007} &159.2(3.4) \cite{Molof1974} &&  && \\
&&&162.6(3) \cite{Safronova1999}&162.7(8) \cite{Ekstrom1995}&&  &&\\
&&&162.71 \cite{Tang2014}&164.7(11.5) \cite{Hall1974}&&  &&\\[2ex] 
&4$S_{1/2}$&3117(20)&3102 \cite{Tang2014}&  &&  & &\\[2ex]
&3$P_{1/2}$&361.0(1.4)&360.05 \cite{Tang2014}&359.6(7) \cite{Windholz1985}&&  &&\\[2ex]   
&3$P_{3/2}$&362.8(1.4)&361.67 \cite{Tang2014}&360.7(8) \cite{Windholz1989}&& $-88.2(2)$&$-88.364$ \cite{Tang2014}&$-88.4(4)$ \cite{Windholz1989}\\[2ex]   
&3$D_{3/2}$&6404.9(4.0)& 6419.5 \cite{Tang2014}&  &&$-3563.3(2.2)$&$-$3566.5 \cite{Tang2014}&\\[2ex]   
&3$D_{5/2}$&6378.8(3.0)&6395.7 \cite{Tang2014}&  &&$-5053.4(1.3)$&$-$5063.7 \cite{Tang2014}&\\[2ex]   
\hline   
K&4$S_{1/2}$&291.9(1.4)&290.4(1.3) \cite{Sahoo2007}, 290.2(8) \cite{Safronova1999}   &292.9(6.1) \cite{Molof1974}&&  &&\\
&&&290.5(1.0) \cite{Nandy2012}, 289.76 \cite{Tang2014} & 305.0(21.6) \cite{Hall1974}&&  &&\\
&&&290.4(6) \cite{Safronova2013}& 290.58(1.42) \cite{Holmgren2010}&&  &&\\[2ex]
&5$S_{1/2}$&4945(39)&4968.7 \cite{Tang2014}, 4961(22) \cite{Safronova2013}& &&  &&\\[2ex]
&4$P_{1/2}$&608.4(7.4)&606(7) \cite{Nandy2012}, 608.9 \cite{Tang2014}&587(88) \cite{Marrus1969}&&  &&\\
&&&612(5) \cite{Safronova2013}& &&  &&\\[2ex]
&4$P_{3/2}$&619.5(7.7)&614(6) \cite{Nandy2012}, 619 \cite{Tang2014}&613(103) \cite{Marrus1969}&&$-109.4(6.0)$&$-106(2)$ \cite{Nandy2012}, $-109.9$ \cite{Tang2014}&$-107(2)$ \cite{Krenn1997}\\
&&&621(4) \cite{Safronova2013}&614(10) \cite{Krenn1997} &&  &$-109.4(1.1)$ \cite{Safronova2013} &\\[2ex]
&3$D_{3/2}$&1417(30)&1466(22) \cite{Nandy2012}, 1426 \cite{Tang2014}&  &&$-480.0(9.7)$&$-$503(13) \cite{Nandy2012}, $-$483.9 \cite{Tang2014}&\\
&&&1420(30) \cite{Safronova2013}& &&  &$-482(19)$ \cite{Safronova2013} &\\[2ex]
&3$D_{5/2}$&1403(26)&1453(33) \cite{Nandy2012}, 1417 \cite{Tang2014}&  &&$-667.6(11.8)$&$-$702(26) \cite{Nandy2012}, $-$767.6 \cite{Tang2014}& \\
&&&1412(30) \cite{Safronova2013}& &&  &$-673(23)$ \cite{Safronova2013} &\\[2ex]
\hline\hline      
\end{tabular}
\label{tab4}
\end{table*}
\section{Results \& Discussion}\label{secre}

In the course of calculating accurate values of polarizabilities, it is necessary to use reliable single-particle orbitals along with considering a powerful many-body method. For our calculations, we have used a large basis set of functions with 40, 39, 38, 37, 36, 35, and 34 Gaussian type orbitals (GTOs) for the $s, p, d, f, g, h$, and $i$ symmetries, respectively, to generate single-particle DHF orbitals. The GTOs are parameterized by two parameters, $\eta_0$ and $\beta$, with the large ($L$) component expressed as $g^{L}_{l,i}=N^Lr^{l+1}e^{-\eta_0\beta^{i-1}}r^2$, for $N^L$ being the normalization constant \cite{Mohanty1989}. The values of $\eta_0$ and $\beta$ used for each symmetry in our calculations are provided in Table \ref{tab_basis}. We present the calculated values for $\alpha_d^S$ and $\alpha_d^T$ of the Li, Na, and K atoms in Table \ref{tab1}. These values were obtained using several methods, including the DHF, MBPT(3), RPA and RCCSD methods. As presented in the table, the differences between the DHF and RPA results are minimal. Since RPA accounts for CP effects to all orders, this small variation indicates that CP effects are not significant in these alkali atoms. To capture contributions from non-RPA effects at intermediate levels, we used the MBPT(3) method, which includes non-RPA effects such as the lowest-order PC contribution. The table clearly highlights that non-RPA effects play a major role in the determination polarizabilities of these atoms. The RCC method, which accounts for both the RPA and non-RPA effects to all-orders, is therefore more accurate than both the RPA and MBPT(3) methods.

Fig. \ref{fig1} shows the trends of $\alpha_d^S$ and $\alpha_d^T$ across different many-body methods, normalized to the corresponding DHF values. The common trend in all these three atoms is that states belonging to the same principal and orbital angular momentum quantum number follows similar behavior. Next, we discuss the correlation trends for each individual atom. As previously discussed, non-RPA effects dominate the polarizability of the Li atom. For scalar polarizability, all Li states exhibit a similar trend, though the $3D_{3/2,5/2}$ states are more sensitive to correlation effects compared to the other states. The non-RPA effects reduce magnitudes of the scalar polarizabilities compared to their DHF values. The only exception to this trend is observed in the tensor polarizability of the $2P_{3/2}$ state, where RPA effects reduce the polarizability and make it negative. However, the dominant non-RPA effects reverse the sign and increase the values overall. For Na, the ground state $3S_{1/2}$ is more sensitive to correlation effects than the other states. Interestingly, the $3D_{3/2,5/2}$ states show an opposite trend in the scalar polarizability results compared to other states. In case of the $3D$ states, non-RPA effects increase the $\alpha_d^S$ value relative to the DHF results. However, for tensor polarizabilities, all states follow the same correlation trends, unlike the trends observed for Li. Similar to Na, the ground state of the K atom, $4S_{1/2}$, is significantly sensitive to the correlation effects compared to the other states. However, unlike Na, all the states in K follow a similar correlation trend for $\alpha_d^S$. For the tensor polarizability, the $4P_{3/2}$ state shows a different trend compared to the $3D$ states. In this case, non-RPA effects lead to an increase in the polarizability value for the $4P_{3/2}$ state. These differences in the trends indicate that although these atoms belong to the same group in the periodic table and share similar chemical properties, correlation behaviors within the systems differ.

To analyze individual contributions, we present the core, core-valence, and valence contributions derived from DHF and different many-body methods in Table \ref{tab2}. The core contribution is minimal compared to the valence contributions, while the core-valence contributions are even less significant. As mentioned earlier, contributions from the closed core are zero for tensor polarizability. In the case of Li, the core and core-valence contribution is negligible. Additionally, Li lacks occupied $P_{1/2,3/2}$ states, resulting in zero core-valence contributions to the polarizability values of the $S_{1/2}$ and $D_{3/2,5/2}$ states. For Na, the core contribution is larger than that of Li, as expected, due to its heavier size. However, the core-valence contribution remains negligibly small. Na possesses occupied $P$ states, thus the core-valence contribution is not zero for the $S$ and $D$ states, unlike in the Li case. In contrast, for the K atom, the core contribution is substantial, and the core-valence contribution for the ground state cannot be disregarded for accurate calculations. Nonetheless, for other excited states, the core-valence contribution becomes negligible again. This trend indicates that as atomic number increases, the core and core-valence contributions also rise, necessitating high-precision calculations to attain accurate polarizability values.

\begin{figure*}
\centering
\includegraphics[width=150mm, height=30mm]{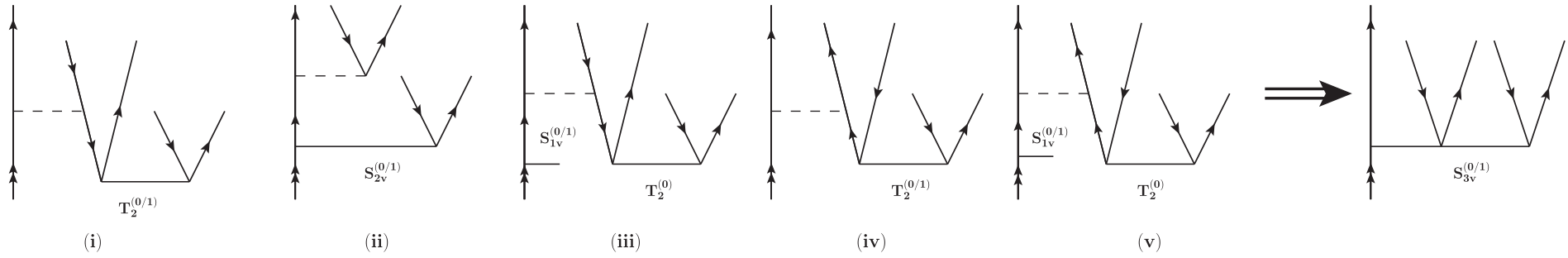}
\caption{Few important Goldstone diagrams that contribute to the unperturbed $S_{3v}^{(0)}$ and perturbed $S_{3v}^{(1)}$ operator. The dotted line corresponds to $V_{res}$ and the line with double arrow represents the valence orbital.}
\label{trip-1}
\end{figure*}

To explore how different correlation effects contribute to the values of $\alpha_d^S$ and $\alpha_d^T$ through the RCC theory, we present results from the individual RCCSD terms in Table \ref{tab3}. The closed part of $\bar{\tilde{D}}T_1^{(1)}$ and its complex conjugate (c.c.) term, $T_1^{(1)\dagger}\bar{\tilde{D}}$, contribute to the core correlation, while the open part contributes to the core-valence correlation. In the table, we have given both core and core-valence contributions separately under the terms $(\bar{\tilde{D}}T_1^{(1)} + c.c.)_{c}$ and $(\bar{\tilde{D}}T_1^{(1)} + c.c.)_{cv}$ respectively. The terms involving $S_v^{(0)}$ or $S_v^{(1)}$ contribute to the valence correlations. From the table, it is evident that the dominant contributions to both $\alpha_d^S$ and $\alpha_d^T$ come from the term $\bar{\tilde{D}}S_{1v}^{(1)}$ and its c.c., followed by $S_{1v}^{(0)\dagger} \bar{\tilde{D}} S_{1v}^{(1)} +$ c.c.. The core correlations arising from $\bar{\tilde{D}} T_1^{(1)}$ and c.c. terms include contributions from both the singly and doubly excited configurations. As a result, these terms account not only for the core contributions of the RPA method but also for the PC contributions to the core correlations of the MBPT(3) method to all-orders. Similarly, the valence correlation contributions from the RPA method are captured by the $\bar{\tilde{D}} S_{1v}^{(1)} + S_{1v}^{(1)\dagger} \bar{\tilde{D}}$ terms in the RCCSD method. These also include contributions from PC correlations and higher-order effects \cite{Sahoo2007, Sahoo2009}. Another important RCCSD term, particularly for the K atom, is $\bar{\tilde{D}} S_{2v}^{(1)} + S_{2v}^{(1)\dagger} \bar{\tilde{D}}$. These contributions cannot be neglected when accurately determining polarizabilities. Additionally, we show corrections to the polarizability values due to wave function normalization, labeled as `Norm'. As shown in Table \ref{tab3}, the contribution from normalization becomes more significant as we move from Li to K, and it is particularly pronounced for the $D$ states. Numerous other correlation contributions to $\alpha_d^S$ and $\alpha_d^T$ also arise from other RCCSD terms, such as $\bar{\tilde{D}}T_{1/2}^{(1)}S_{1/2v}^{(0)}$, $T_{1/2}^{(1)\dagger} \bar{\tilde{D}} S_{1/2v}^{(0)}$, $S_{1/2v}^{(0)\dagger} \bar{\tilde{D}} S_{1/2v}^{(1)}$ and $S_{1/2v}^{(1)\dagger} \bar{\tilde{D}} S_{1/2v}^{(0)}$. These are non-RPA effects, many of which cannot be considered part of the PC correlation. We present the contributions from these nonlinear terms under the label `others' in the table above.

The recommended values for $\alpha_d^S$ and $\alpha_d^T$ from the RCCSD method are presented in Table \ref{tab4}, along with the estimated uncertainties, which are derived from the leading order triple excitations. Several of the dominantly contributing triple excitation diagrams are depicted in Fig. \ref{trip-1}. We have calculated the polarizability values by incorporating contributions from these triple excitations perturbatively and estimated the uncertainty by comparing the results with our RCCSD calculations. Numerous studies have reported calculated and experimental values for the polarizability of these alkali atoms \cite{Wansbeek2008, Molof1974, Ekstrom1995, Hall1974, Miffre2006, Safronova2012, Safronova1999, Nandy2012, Marrus1969, Tang2010, Tang2014, Themelis1995, Windholz1992, Hunter1991, Johnson2008, Rerat1998, Windholz1985, Windholz1989, Holmgren2010, Safronova2008, Safronova2013, Arora2007, Schmieder1971, Krenn1997}. In the table, we compare our results with recent high-precision relativistic calculations and experimental data. For the Li atom, our {\it ab initio} results show good agreement with the experimental values for the $2S_{1/2}$ and $2P_{1/2,3/2}$ states \cite{Molof1974, Miffre2006, Windholz1992, Hunter1991}. Since no experimental values are available for the other excited states of Li, we compare our results with previously reported theoretical values \cite{Wansbeek2008, Safronova2012, Tang2014}. As shown in the table, our result for the $3S_{1/2}$ state is consistent with previous calculations. However, for the $3D_{3/2,5/2}$ states, there is a discrepancy of about 1\% with earlier reported results. Notably, Ref. \cite{Wansbeek2008} used a hybrid approach of combining numerical and analytical basis functions, while Ref. \cite{Safronova2012} applied a sum-over-states method using a mix of experimental and theoretical energy values. In Ref. \cite{Tang2014}, a semi-empirical DHF $+$ CP potential approach was adopted to estimate polarizabilities. In contrast, our work employs a purely {\it ab initio} approach. The key reason for this discrepancy is the use of the sum-over-states approach in the earlier studies. While this method relies on experimental energies for low-lying valence states, it is not applicable to core orbitals or high-lying valence and continuum states. In those earlier calculations, contributions from these orbitals were estimated using the mean-field DHF method, or RPA, which led to the omission of several important correlation contributions. In contrast, the {\it ab initio} RCCSD method treats all orbitals on an equal footing. This causes inconsistencies with the sum-over-states approach, resulting in discrepancies in the polarizability values. For the Na atom, our polarizability results for the ground state and the $3P_{1/2,3/2}$ states are in good agreement with the available experimental values \cite{Molof1974, Ekstrom1995, Hall1974, Windholz1985}. For other excited states, we compare our results with Ref. \cite{Tang2014}. Our recommended results for the polarizability values of the $3D_{3/2,5/2}$ states are in agreement with that of Ref. \cite{Tang2014}. For Na, non-relativistic results are also available \cite{Themelis1995, Rerat1998}. However, significant discrepancies exist between our results and those reported in previous studies, primarily due to the non-relativistic approach employed in those calculations. We also compare our results for the K atom with the available experimental and theoretical values in the same table. As presented in the table, our ground state polarizability value agrees well with the experimental results \cite{Molof1974, Hall1974, Holmgren2010}. For the $4P_{1/2,3/2}$ states, the error bar in the experimental results for $\alpha_d^S$ is relatively large \cite{Marrus1969}, while the error bar in our calculation is much smaller. Nonetheless, our results align with other theoretical values \cite{Nandy2012,  Safronova2013, Tang2014}. For the $3D_{3/2,5/2}$ states, while no experimental values are available, our calculations fall within the error bars of other relativistic sum-over-states calculations \cite{Nandy2012, Safronova2013}. Although there are non-relativistic results for the $3D$ states available, they do not match our recommended values \cite{Rerat1998}. Similarly, our tensor polarizability values are consistent with the error bars of Refs. \cite{Nandy2012, Safronova2013}.

\section{Summary}

We have employed linear response coupled-cluster theory in the relativistic framework to determine the scalar and tensor polarizabilities of both the ground and several excited states of Li, Na and K atom. By considering the singles- and doubles-excitation approximation and estimating uncertainties from the neglected triple contributions, accurate {\it ab initio} values for the above quantities are reported. We have also given values from random phase approximation and the third-order perturbation theory from our calculations to understand propagation of correlation effects from lower- to higher-order perturbation theory. It was found that the core polarization effects are not prominent, but the pair-correlations and relativistic effects play important roles in the determination of electric dipole polarizabilities in the considered alkali atoms. We also compared our results with the available experimental values and observed that our results from the relativistic coupled-cluster theory are in good agreement with them.

\section*{Acknowledgment}

This work is supported by ANRF with grant no. CRG/2023/002558 and Department of Space, Government of India. Calculations were carried out using the ParamVikram-1000 HPC cluster of the Physical Research Laboratory (PRL), Ahmedabad, Gujarat, India.


\begin{thebibliography}{5}
\bibitem{Bonin1997}
K. D. Bonin and V. V. Kresin, {\it Electric Dipole Polarizabilities of Atoms, Molecules and Clusters}, World Scientific, Signapore (1997).

\bibitem{Haken1996}
H. Haken and H. C. Wolf, {\it Atoms in an Electric Field. In: The Physics of Atoms and Quanta}, Springer (Berlin, Heidelberg), pp. 251-279 (1996). 

\bibitem{Manakov1986}
N. L. Manakov and V. D. Ovsiannikov and L. P. Rapoport, Phys. Rep. {\bf 141}, 320 (1986).

\bibitem{Buckingham1967}
A. D. Buckingham, Adv. Chem. Phys. {\bf 12}, 107 (1967).

\bibitem{Maroulis2004}
G. Maroulis, George, J. Comput. Methods Sci. Eng. {\bf 4}, 235 (2004). 

\bibitem{Pal'chikov2003}
V G Pal'chikov, Yu S Domnin and A V Novoselov, J. Opt. B: Quantum Semiclass. Opt. {\bf 5}, S131 (2003).

\bibitem{Sahoo2014}
B. K. Sahoo, Pramana {\bf 83}, 255 (2014).

\bibitem{Safronova2012-IEEE}
M. S. Safronova, M. G. Kozlov, and C. W. clark, IEEE Trans. on Ult. Ferro. and Freq. cntrl. , {\bf 59}, 439 (2012). 

\bibitem{Burrow1976}
P. Burrow, J. Michejda, J. Comer, J. Phys. B At. Mol. Phys. {\bf 9}, 3225 (1976).

\bibitem{Jain1990}
A. Jain, Phys. Rev. A {\bf 41}, 2437 (1990).

\bibitem{Tenfen2019}
W. Tenfen, M. V. Barp, and F. Arretche, Phys. Rev. A {\bf 99}, 022703 (2019).

\bibitem{Tang1969}
K. T. Tang, Phys. Rev. {\bf 177}, 108 (1969).

\bibitem{Derevianko1999}
A. Derevianko, W. R. Johnson, M. S. Safronova, and J. F. Babb, Phys. Rev. Lett. {\bf 82}, 3589 (1999). 

\bibitem{Wansbeek2008}
L. W. Wansbeek, B. K. Sahoo, R. G. E. Timmermans, B. P. Das, and D. Mukherjee, Phys. Rev. A {\bf 78}, 012515 (2008).

\bibitem{Arora2014}
B. Arora, and B. K. Sahoo, Phys. Rev. A {\bf 89}, 022511 (2014).

\bibitem{Monin1974}
J. Monin and G. -A. Boutry, Phys. Rev. B {\bf 9}, 1309 (1974).

\bibitem{Sahoo2007}
B. K. Sahoo, Chem. Phys. Lett. {\bf 448}, 144 (2007).

\bibitem{Sahoo2009}
B. K. Sahoo and B. P. Das, and D. Mukherjee, Phys. Rev. A {\bf 79}, 052511 (2009).

\bibitem{Oymak2012}
H. Oymak and S. Erkoc, Int. J. Mod. Phys. B {\bf 26},
1230003 (2012). 

\bibitem{Peter2019}
P. Schwerdtfeger and J. K. Nagle, Mol. Phys. {\bf 117}, 1200  (2019).

\bibitem{Ravi2012}
K. Ravi, S. Lee, A. Sharma, G. Werth, and S. Rangwala, Nat. Comm. {\bf 3}, 1126 (2012).

\bibitem{Hall2012}
F. H. J. Hall and S. Willitsch, Phys. Rev. Lett. {\bf 109}, 233202 (2012).

\bibitem{Catani2006}
J. Catani, P. Maioli, L. De Sarlo, F. Minardi, and M. Inguscio, Phys. Rev. A {\bf 73}, 033415 (2006).

\bibitem{Saffman2010}
M. Saffman, T. G. Walker, and K. Mølmer, Rev. Mod. Phys. {\bf 82}, 2313 (2010).

\bibitem{Joachim2000}
C. Joachim, J. K. Gimzewski, and A. Aviram, Nature (London) {\bf 408}, 541 (2000).

\bibitem{Molof1974} 
R. W. Molof, H.L. Schwartz, T.M. Miller, B. Bederson, Phys. Rev. A {\bf 10}, 1131 (1974).

\bibitem{Ekstrom1995}
C. R. Ekstrom, J. Schmiedmayer, M.S. Chapman, T.D. Hammond, and D.E. Pritchard, Phys. Rev. A {\bf 51}, 3883 (1995).


\bibitem{Hall1974}
W. D. Hall and J. C. Zorn, Phys. Rev. A {\bf 10}, 1141 (1974).

\bibitem{Miffre2006}
A. Miffre, M. Jacquey, M. Büchner, G. Trenec, and J. Vigue, Phys. Rev. A {\bf 73}, 011603(R) (2006).

\bibitem{Safronova2012}
M. S. Safronova, U. I. Safronova, and Charles W. Clark, Phys. Rev. A {\bf 86}, 042505 (2012).

\bibitem{Safronova1999}
M. S. Safronova, W. R. Johnson, and A. Derevianko, Phys. Rev. A {\bf 60}, 4476 (1999).

\bibitem{Nandy2012}
D. K. Nandy, Y. Singh, B. P. Shah, and B. K. Sahoo, Phys. Rev. A {\bf 86}, 052517 (2012).

\bibitem{Marrus1969}
R. Marrus and J. Yellin, Phys. Rev. {\bf 177}, 127 (1969).

\bibitem{Dalgarno1955}
A. Dalgarno and J. T. Lewis, Proc. R. Soc. London {\bf 233}, 70 (1955).

\bibitem{Sahoo2025}
B. K. Sahoo, S. Blundell, A. V. Oleynichenko, R. F. Garcia Ruiz, L. V. Skripnikov and B. Ohayon, J. Phys. B {\bf 58}, 042001 (Topical review) (2025).

\bibitem{Katyal2025}
V. Katyal, A. Chakraborty, B. K. Sahoo, B. Ohayon, C.-Y. Seng, M. Gorchtein and J. Behr,  	arXiv:2412.05932 (Unpublished).

\bibitem{Kozlov1999}
M. G. Kozlov and S. G. Porsev, Eur. Phys. J. D {\bf 5}, 59 (1999).

\bibitem{Stalnaker2006}
J. E. Stalnaker, D. Budker, S. J. Freedman, J. S. Guzman, S. M. Rochester, and V. V. Yashchuk, Phys. Rev. A {\bf 73}, 043416 (2006).

\bibitem{Arora2012}
B. Arora, D. K. Nandy, and B. K. Sahoo, Phys. Rev. A {\bf 85}, 012506 (2012).

\bibitem{Kaur2015}
J. Kaur, D. K. Nandy, B. Arora, and B. K. Sahoo, Phys. Rev. A 91, 012705 (2015).


\bibitem{Lindgren1985}
I. Lindgren and J. Morrison, in {\it Atomic Many-Body Theory}, edited by G. Ecker, P. Lambropoulos, and H. Walther (SpringerVerlag, Berlin, 1985).

\bibitem{Singh2014}
Y. Singh, B. K. Sahoo, and B. P. Das, Phys. Rev. A {\bf 89}, 030502(R) (2014).

\bibitem{Chakraborty2023-2}
A. Chakraborty and B. K. Sahoo, J. Phys. Chem. A  {\bf 127}, 7518 (2023).

\bibitem{Sahoo2017}
B. K. Sahoo, Y. Singh, Phys. Rev. A {\bf 95}, 062514 (2017).



\bibitem{Mukherjee1979}
D. Mukherjee, Pramana {\bf 12}, 203 (1979).

\bibitem{Mohanty1989}
A. K. Mohanty and E. Clementi, Chem. Phys. Lett. {\bf 157}, 348 (1989).


\bibitem{Tang2010}
Li-Yan Tang, Zong-Chao Yan, Ting-Yun Shi, and J. Mitroy, Phys. Rev. A {\bf 81}, 042521 (2010).

\bibitem{Tang2014}
Tang Yong-Bo, Li Cheng-Bin and Qiao Hao-Xue, Chin. Phys. B {\bf 23}, 063101 (2014).


\bibitem{Themelis1995}
S. I. Themelis and C. A. Nicolaides, Phys. Rev. A 51, 2801 (1995).

\bibitem{Windholz1992}
L. Windholz, M. Musso, G. Zerza, and H. Jager, Phys. Rev. A
{\bf 46}, 5812 (1992).

\bibitem{Hunter1991}
L. R. Hunter, D. Krause, D. J. Berkeland, and M. G. Boshier,
Phys. Rev. A {\bf 44}, 6140 (1991).

\bibitem{Johnson2008}
W. R. Johnson, U. I. Safronova, A. Derevianko, and M. S.
Safronova, Phys. Rev. A 77, 022510 (2008).

\bibitem{Rerat1998}
M. Rerat, M. Merawa, and B. Honvault-Bussery, J. Chem. Phys. {\bf 109}, 7246 (1998).

\bibitem{Windholz1985}
L. Windholz and C. Neureiter, Phys Lett. {\bf 109A}, 155 (1985).    

\bibitem{Windholz1989}
L. Windholz and M. Musso, Phys Rev. A {\bf 39}, 2472 (1989).

\bibitem{Holmgren2010}
W. F. Holmgren, M. C. Revelle, V. P. A. Lonij, and A. D. Cronin, Phys. Rev. A {\bf 81}, 053607 (2010).

\bibitem{Safronova2008}
U. I. Safronova and M. S. Safronova, Phys. Rev. A  {\bf 78}, 052504 (2008). 

\bibitem{Safronova2013}
M. S. Safronova, U. I. Safronova and C. W. Clark, Phys. Rev. A {\bf 87}, 052504 (2013).

\bibitem{Arora2007}
B. Arora, M. S. Safronova, and C. W. Clark, Phys. Rev. A {\bf 76}, 052509 (2007).

\bibitem{Schmieder1971}
R. Schmieder, A. Lurio, and W. Happer, Phys. Rev. A {\bf 3}, 1209 (1971).

\bibitem{Krenn1997}
C. Krenn, W. Scherf, O. Khait, M. Musso, and L. Windholz, Z. Phys. D {\bf 41}, 229 (1997).



\end{thebibliography}
\end{document}